\documentclass[a4paper,11pt]{article}
\pdfoutput=1 

\usepackage{jheppub} 

\usepackage[T1]{fontenc}
\usepackage{multirow}

\allowdisplaybreaks
 
\definecolor{nicered}{rgb}{0.7,0.1,0.1}
\definecolor{nicegreen}{rgb}{0.1,0.5,0.1}
\definecolor{niceblue}{rgb}{0.0,0.1,0.7}
\hypersetup{colorlinks,citecolor=niceblue,linkcolor=niceblue,urlcolor=niceblue}

\usepackage[normalem]{ulem}

\def \beq{\begin{equation}}
\def \eeq{\end{equation}}
\def \bea{\begin{eqnarray}}
\def \eea{\end{eqnarray}}

\preprint{MPP-2025-104}

\title{A new probe of the quartic Higgs self-coupling}

\author[a]{Ulrich Haisch,}
\author[a,b]{Aparna Sankar}
\author[a,b]{and Giulia Zanderighi}

\affiliation[a]{Max Planck Institute for Physics, \\ Boltzmannstr.~8, 85748 Garching, Germany}
\affiliation[b]{Technische Universit{\"a}t M{\"u}nchen, Physik-Department, \\ James-Franck-Straße 1, 85748 Garching, Germany }

\emailAdd{haisch@mpp.mpg.de}
\emailAdd{aparna@mpp.mpg.de}
\emailAdd{zanderi@mpp.mpg.de}

\abstract{We calculate the corrections to the Higgs wave-function renormalization constant arising from modified cubic, quartic, and quintic Higgs self-couplings up to the two-loop level. Using our analytic results, we derive two-dimensional constraints on the modifications of the considered Higgs self-interactions that could potentially be set from precision measurements of single-Higgs production processes at the high-luminosity Large Hadron Collider~(LHC) and a Future Circular Collider. Our novel constraints are compared to those that might be set by searches for multi-Higgs production at the same~facilities. In view of the first LHC results on triple-Higgs production, we also review the current status of Higgs self-coupling determinations after LHC~Run~2.}

\begin{document} 
\maketitle
\flushbottom

\section{Introduction}
\label{sec:introduction}

After electroweak~(EW) symmetry breaking, the cubic, quartic, and quintic self-interactions of the Higgs field $h$ can be parameterized in a model-independent fashion by 
\beq \label{eq:V}
V \supset \kappa_3 \hspace{0.25mm} \lambda \hspace{0.25mm} v \hspace{0.25mm} h^3 + \kappa_4 \hspace{0.25mm} \frac{\lambda}{4} \hspace{0.25mm} h^4 + \kappa_5 \hspace{0.25mm} \frac{\lambda}{v} \hspace{0.25mm} h^5 \,.
\eeq
Here, $\lambda = m_h^2/(2 \hspace{0.25mm} v^2)$ with $m_h \simeq 125 \, {\rm GeV}$ being the Higgs mass and $v \simeq 246 \, {\rm GeV}$ being the EW vacuum expectation value. Notice that the normalization of the terms in~(\ref{eq:V}) has been chosen such that within the Standard Model~(SM) one has~$\kappa_{3} = \kappa_{4} = 1$, while $\kappa_5 = 0$. 

In the presence of physics beyond the SM (BSM), the coefficients $\kappa_3$, $\kappa_4$, and $\kappa_5$ will generally deviate from their respective SM values. As an illustrative example, let us consider the following two terms
\beq \label{eq:LSMEFT}
{\cal L}_{\rm SMEFT}\supset Q_6 + Q_8 = -\frac{\lambda \hspace{0.25mm} \bar c_6}{v^2} \hspace{0.25mm} \left |H \right |^6 -\frac{\lambda \hspace{0.25mm} \bar c_8}{v^4} \hspace{0.25mm} \left |H \right |^8 \,,
\eeq
in the SM effective field theory (SMEFT)~\cite{Buchmuller:1985jz,Grzadkowski:2010es,Brivio:2017vri,Isidori:2023pyp}, where $H$ denotes the usual Higgs doublet. In this case, the parameters $\kappa_3$, $\kappa_4$, and $\kappa_5$ are related to the Wilson coefficients $\bar c_6$ and $\bar c_8$. At~tree~level, the relationships are given by:
\beq \label{eq:kappa3kappa4kappa5}
\kappa_3 = 1 + \bar c_6 + 2 \hspace{0.25mm} \bar c_8 \,, \qquad 
\kappa_4 = 1 + 6 \hspace{0.25mm} \bar c_6 + 16 \hspace{0.25mm} \bar c_8 \,, \qquad 
\kappa_5 = \frac{3}{4} \hspace{0.5mm} \bar c_6 + \frac{7}{2} \hspace{0.5mm} \bar c_8 \,.
\eeq
The relations~(\ref{eq:kappa3kappa4kappa5}) imply that if the dimension-six operator $Q_6$ is the only numerically significant modification in the SMEFT, the shifts in the cubic, quartic, and quintic Higgs self-couplings are strongly correlated. For instance, one has $\kappa_4 - 1 = 6 \left ( \kappa_3 - 1 \right )$. However, this correlation is broken by the dimension-eight contribution $Q_8$, if this operator has a non-zero Wilson coefficient. The initial conditions of the Wilson coefficients~$\bar c_6$ and~$\bar c_8$ can be determined in any ultraviolet~(UV) complete BSM model through a suitable matching calculation. If the new interactions leading to $Q_6$ and $Q_8$ are weakly coupled and the new-physics scale $\Lambda$ is in the TeV range, one generally expects the dimension-eight and dimension-six contributions to follow the hierarchy~$\bar c_8/\bar c_6 = {\cal O} (v^2/\Lambda^2) \ll 1$. However, the Wilson coefficients~$\bar c_8$ and~$\bar c_6$ can be of the same order of magnitude if the underlying UV theory is strongly coupled or if the new-physics scale $\Lambda$ is at or not far above the EW scale. To~achieve the inverted hierarchy $\bar c_8/\bar c_6 \gg 1$ the new particles that give rise to~(\ref{eq:LSMEFT}) must have masses of ${\cal O} (v)$ and strong interactions with the Higgs field $H$. SM~extensions with colorless $SU(2)$ quadruplets~$\Theta$~\cite{deBlas:2014mba} can, for instance, lead to such an inverted hierarchy if the quadruplet is sufficiently light and the Higgs portal coupling $|\Theta|^2 \, |H|^2$ is large enough. See~also~\cite{Durieux:2022hbu} for a general discussion of UV~models where BSM physics dominantly manifest itself in Higgs self-coupling modifications. Notice that, in order to decorrelate $\kappa_5$ from $\kappa_3$ and $\kappa_4$, one can, for example, add the operator $Q_{10} = -\lambda \hspace{0.25mm} \bar c_{10}/v^6 \hspace{0.25mm} \left |H \right |^{10}$ to the Lagrangian~(\ref{eq:LSMEFT}). Similar arguments to those made for $Q_8$ then also apply to $Q_{10}$. 	

In our article, we take an agnostic approach to how new UV dynamics affect the Higgs self-interactions, making no assumptions about the actual magnitudes of the Wilson coefficients of the pure Higgs SMEFT operators. The cubic, quartic, and quintic Higgs self-couplings can then deviate independently from the SM predictions. The important point is that even if $\kappa_{3}$, $\kappa_{4}$, and $\kappa_{5}$ are treated as free parameters, loop corrections to $e^+ e^- \to hZ$ or $gg \to h$ can still be calculated consistently, as long as the SMEFT is used to perform the computations~(see~\cite{McCullough:2013rea,Gorbahn:2016uoy,Degrassi:2016wml,Bizon:2016wgr,Degrassi:2017ucl,Kribs:2017znd,DiVita:2017eyz,Maltoni:2017ims,DiVita:2017vrr,Maltoni:2018ttu,Liu:2018peg,Bizon:2018syu,Borowka:2018pxx,Gorbahn:2019lwq,Degrassi:2019yix,Haisch:2021hvy,Gao:2023bll,Li:2024iio,Haisch:2024nzv} for non-trivial one-loop and two-loop examples and further explanations). Upon canonically normalizing the Higgs field, a universal correction to all on-shell Higgs observables arises from Higgs wave-function renormalization (WFR). This rescaling is important because, while it can be adjusted by rescaling other fields or couplings, it cannot be eliminated entirely from the theory. In fact, it generally disrupts the SM prediction for the relationship between the mass of a particle and its coupling to the Higgs, making it observable~\cite{Englert:2013tya,Craig:2013xia}. Indeed, it has been demonstrated in~\cite{McCullough:2013rea,Gorbahn:2016uoy,Degrassi:2016wml,Bizon:2016wgr,Degrassi:2017ucl,Gorbahn:2019lwq,Degrassi:2019yix} that in single-Higgs production and decay processes, the dominant contributions arise from the one-loop Higgs~WFR constant, which exhibits a quadratic dependence on $\kappa_3$. This dominance can be attributed to the currently weak experimental constraints (cf.~for~example~\cite{ATLAS:2022jtk,CMS:2024awa}) on possible modifications of the cubic Higgs self-coupling.

While constraints on $\kappa_3 $ are anticipated to significantly improve by leveraging data on double-Higgs production at the high-luminosity Large Hadron Collider (HL-LHC), bounding the quartic Higgs self-coupling will remain notoriously difficult at both future hadron and lepton colliders~\cite{Maltoni:2018ttu,Liu:2018peg,Bizon:2018syu,Borowka:2018pxx,Chiesa:2020awd,Gonzalez-Lopez:2020lpd,Stylianou:2023xit,Papaefstathiou:2023uum,Brigljevic:2024vuv,Dong:2025lkm}. In this context, it is important to explore as many complementary approaches as possible to constrain $\kappa_4$. The goal of this article is to propose and develop a new method for probing the quartic Higgs self-coupling. It is based on a two-loop calculation of the Higgs~WFR constant, retaining the full dependence on $\kappa_3$, $\kappa_4$, and~$\kappa_5$. This~calculation is described in~Section~\ref{sec:calculation}. In Section~\ref{sec:numerics}, we then determine the hypothetical reach of the HL-LHC and a Future Circular Collider (FCC) in constraining the relevant Higgs self-couplings through measurements of single-Higgs production processes. We compare our constraints to those that searches for multi-Higgs production might set in the future at the same facilities. Section~\ref{sec:conclusions} contains our conclusions. Additional material is provided in four appendices. In~Appendix~\ref{app:integrals}, we present the analytic results for all the master integrals involved in the two-loop computation of the pure Higgs contribution to the Higgs self-energy. In~Appendix~\ref{app:counterterms}, we include the Feynman rules for the counterterms used in our computation. The~renormalization scale dependence of the single-Higgs production constraints at the HL-LHC and the FCC is discussed in~Appendix~\ref{app:moresingle}. In light of the recent ATLAS results on triple-Higgs production~\cite{ATLAS:2024xcs}, we conclude our article by discussing the current status of Higgs self-coupling extractions after LHC~Run~2 in Appendix~\ref{app:LHCRun2}.

\section{Calculation}
\label{sec:calculation}

All calculations in this article use dimensional regularization in~$d = 4 -2 \hspace{0.125mm} \epsilon$ dimensions, with~$\epsilon > 0$, supplemented by the renormalisation scale~$\mu$. The generation and computation of the amplitudes were carried out using the \texttt{Mathematica} packages \texttt{FeynArts}~\cite{Hahn:2000kx}, \texttt{FeynCalc}~\cite{Shtabovenko:2020gxv}, and \texttt{FormCalc}~\cite{Hahn:2016ebn}. The one-loop and two-loop integrals were reduced using the Tarasov algorithm~\cite{Tarasov:1997kx} to the set of scalar master integrals provided in~Appendix~\ref{app:integrals}. This~algorithm is implemented in the program \texttt{TARCER}~\cite{Mertig:1998vk}, which is part of \texttt{FeynCalc}. The~\texttt{TARCER} reduction has been cross-checked with \texttt{LiteRed}~\cite{Lee:2013mka}, resulting in the same final~outcome.

The perturbative expansion of the pure Higgs contribution to the Higgs~WFR constant can be written as 
\beq \label{eq:Zh}
Z_h = 1 + \frac{\lambda}{(4 \pi)^2} \hspace{0.25mm} \delta Z_h^{(1)} + \left ( \frac{\lambda}{(4 \pi)^2} \right )^2 \hspace{0.25mm} \delta Z_h^{(2)} \,,
\eeq
where $\delta Z_h^{(1)}$ and $\delta Z_h^{(2)}$ denote the one-loop and two-loop corrections, respectively. In the on-shell scheme, the one-loop Higgs~WFR constant is obtained by calculating the first derivative of the unrenormalized Higgs self-energy $\Sigma^{(1)} (p^2)$ with respect to $p^2$, evaluated on-shell:
\beq \label{eq:onshelldZ1}
\frac{\lambda}{(4 \pi)^2} \hspace{0.25mm} \delta Z_h^{(1)} = \left. -\frac{\partial \Sigma^{(1)} (p^2)}{\partial p^2} \right |_{p^2 = m_h^2} = - \Sigma^{\prime \hspace{0.25mm} (1)} (m_h^2) \,. 
\eeq
Here, $p$ denotes the external four-momentum flowing through the Higgs self-energy graphs. A~straightforward one-loop computation leads to the well-known result~\cite{Gorbahn:2016uoy,Degrassi:2016wml}
\beq \label{eq:dZ1}
\delta Z_h^{(1)} = c_{2,0}^{(1)} \hspace{0.5mm} \kappa_3^2 \,, \qquad c_{2,0}^{(1)} = 9 - 2 \sqrt{3} \pi \simeq -1.8828 \,.
\eeq

\begin{figure}[t!]
\begin{center}
\includegraphics[width=0.95\textwidth]{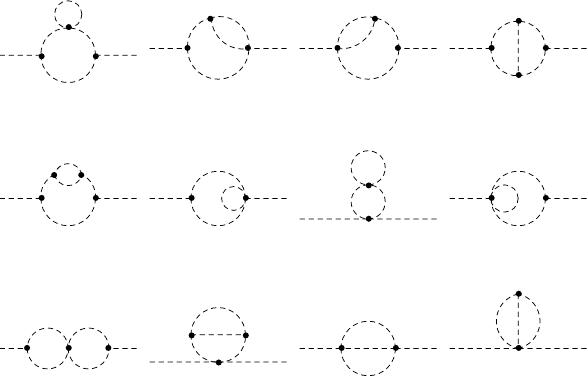}
\end{center}
\vspace{0mm} 
\caption{\label{fig:2loop} 1PI two-loop contributions to the Higgs self-energy. The~dashed lines represent Higgs propagators.}
\end{figure}

\begin{figure}[t!]
\begin{center}
\includegraphics[width=0.75\textwidth]{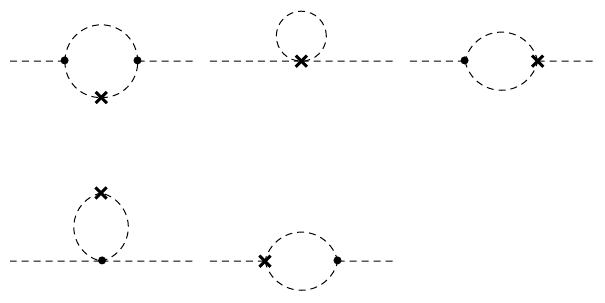}
\end{center}
\vspace{0mm} 
\caption{\label{fig:1loopct} 1PI one-loop counterterm contributions to the Higgs self-energy. The black crosses indicate the counterterm insertions.}
\end{figure}

The two-loop correction to the on-shell Higgs~WFR constant is computed similarly to the one-loop case, using the unrenormalized two-loop Higgs self-energy $\Sigma^{(2)} (p^2)$ and the corresponding one-loop counterterm $\Sigma^{(2)}_{\rm ct} (p^2)$: 
\beq \label{eq:onshelldZ2}
\left ( \frac{\lambda}{(4 \pi)^2} \right )^2 \hspace{0.25mm} \delta Z_h^{(2)} = -\Sigma^{\prime \hspace{0.25mm} (2)} (m_h^2) - \Sigma^{\prime \hspace{0.25mm} (2)}_{\rm ct} (m_h^2) \,. 
\eeq
The two-loop and the one-loop counterterm contributions to the Higgs self-energy involving Higgs exchange are shown in~Figure~\ref{fig:2loop} and Figure~\ref{fig:1loopct}, respectively. Notice that only one-particle-irreducible (1PI) diagrams are considered. In particular, two-loop diagrams involving one-loop Higgs tadpoles are excluded since such contributions are exactly canceled by the corresponding one-loop counterterm contributions. This is because we use the standard renormalization of the Higgs tadpole~\cite{Denner:1991kt}, which sets the renormalized Higgs tadpole to zero at each order of perturbation theory. This choice ensures that the effective potential contains no term linear in the Higgs field. In addition to renormalizing the Higgs tadpole, we also account for Higgs mass and WFR, as well as operator mixing in the SMEFT. The~explicit form of the counterterms used in our calculation is provided in~Appendix~\ref{app:counterterms}. 

Our final result for the two-loop correction to the on-shell Higgs~WFR constant reads 
\beq \label{eq:dZ2}
\begin{split}
\delta Z_h^{(2)} & = - \left [ \frac{3}{2 \hspace{0.25mm} \bar \epsilon} - \frac{9}{4} + 3 L \right ] \kappa_4^2 + \frac{1}{3} \hspace{0.5mm} c_{2,0}^{(1)} \hspace{0.5mm} \big ( 1 + L \big ) \hspace{0.5mm} \kappa_3 \hspace{0.25mm} \kappa_5 - 6 \hspace{0.25mm} c_{2,0}^{(1)} \hspace{0.5mm} \big ( 1 + L \big ) \hspace{0.5mm} \kappa_3^3 \\[2mm]
& \phantom{xx} + \left [ c_{2,1}^{(2)} - 12 \hspace{0.25mm} c_{2,0}^{(1)} \hspace{0.25mm} L \right ] \kappa_3^2 \hspace{0.25mm} \kappa_4 + \left [ c_{4,0}^{(2)} + 18 \hspace{0.25mm} c_{2,0}^{(1)} \hspace{0.25mm} L \right ] \kappa_3^4 \,
\end{split}
\eeq
with $L = \ln \left ( \mu^2/m_h^2 \right )$ and
\beq 
\begin{split} \label{eq:c2loop}
c_{2,1}^{(2)} & = -378 - 36 \hspace{0.25mm} \zeta(2) + 78 \sqrt{3} \pi \simeq -12.7886 \,, \\[4mm]
c_{4,0}^{(2)} & = 405 + 189 \hspace{0.25mm} \zeta(2) - 162 \hspace{0.25mm} \zeta(3) -216 \sqrt{3} \pi + 54 \left ( 3 \sqrt{3} + 2 \pi \right) {\rm Cl}_2 \left(\frac{\pi}{3}\right) \simeq -25.0364 \,. 
\end{split} 
\eeq
Here, we introduced the shorthand notation $1/\bar \epsilon = 1/\epsilon - \gamma_E + \ln \left ( 4 \pi \right)$, where $\gamma_E \simeq 0.57722$ is the Euler–Mascheroni constant. The normalization of our loop integrals --- see~(\ref{eq:MIsdef}), (\ref{eq:MIs1}), and (\ref{eq:MIs2}) --- includes the factor $\left(4\pi\right)^\epsilon e^{-\gamma_E \epsilon}$, which expands as $1 - \gamma_E \hspace{0.25mm} \epsilon + \ln \left (4\pi \right )\epsilon + \mathcal{O}(\epsilon^2)$. As~a result, the pure $1/\epsilon$ poles naturally appear together with the $\gamma_E$ and $\ln \left (4\pi \right )$ terms. Additionally, we note that $\zeta(2) = \pi^2/6 \simeq 1.64493$ and $\zeta(3) \simeq 1.20206$ correspond to the Riemann zeta function evaluated at 2 and 3, respectively, while ${\rm Cl}_2 \left(\pi/3\right) \simeq 1.01494$ denotes the Clausen function at $\pi/3$. Interestingly, the terms $\pi/\sqrt{3} \hspace{0.25mm} \ln 3$ that appear in the master integrals given in~(\ref{eq:MIs1}) and~(\ref{eq:MIs2}) cancel out in the coefficients $c_{2,1}^{(2)}$ and~$c_{4,0}^{(2)}$. The~expressions in (\ref{eq:dZ2}) and (\ref{eq:c2loop}) constitute the main results of this article. Notably, as in the case of double-Higgs production studied in~\cite{Maltoni:2018ttu,Liu:2018peg,Bizon:2018syu,Borowka:2018pxx}, the expression~(\ref{eq:dZ2}) for the two-loop correction to the on-shell Higgs WFR constant depends on the cubic, quartic, and quintic Higgs self-couplings, parameterized by $\kappa_3$, $\kappa_4$, and $\kappa_5$, respectively.

It is worth mentioning that the results in~(\ref{eq:dZ1}) and~(\ref{eq:dZ2}) can be used to calculate the on-shell Higgs~WFR constant in the SM up to the two-loop level. By setting $\kappa_3 = \kappa_4 = 1$ and $\kappa_5 = 0$, we obtain the SM results: 
\beq \label{eq:dZSM}
\Big ( \delta Z_h^{(1)} \Big )_{\rm SM} = c_{2,0}^{(1)} \,, \qquad 
\Big ( \delta Z_h^{(2)} \Big)_{\rm SM} = - \frac{3}{2 \hspace{0.25mm} \bar \epsilon} + \frac{9}{4} - 3 L - 6 \hspace{0.25mm} c_{2,0}^{(1)} + c_{2,1}^{(2)} + c_{4,0}^{(2)} \,, 
\eeq
with the coefficients $c_{2,0}^{(1)}$, $c_{2,1}^{(2)}$, and $c_{4,0}^{(2)}$ given in~(\ref{eq:dZ1}) and~(\ref{eq:c2loop}). Notice that the logarithms in~(\ref{eq:dZ2}) proportional to $\kappa_3 \hspace{0.25mm} \kappa_5$, $\kappa_3^3$, $\kappa_3^2 \hspace{0.25mm} \kappa_4$, and $\kappa_3^4$ cancel in $\Big ( \delta Z_h^{(2)} \Big)_{\rm SM}$, and as a result the term proportional to $L$ in~(\ref{eq:dZSM}) arises solely from the $\kappa_4^2$ contribution to~(\ref{eq:dZ2}). In fact, the two-loop Higgs anomalous dimension that follows from~(\ref{eq:dZSM}) matches the known two-loop SM value calculated, for instance, in~\cite{Machacek:1983tz,Luo:2002ti}, serving as a cross-check of our calculation. Finally, we note that the two-loop Higgs anomalous dimension can be easily obtained by calculating the UV pole of the two-loop sunrise diagram, i.e.,~the next-to-last graph in~Figure~\ref{fig:2loop}.

\begin{figure}[t!]
\begin{center}
\includegraphics[width=0.75\textwidth]{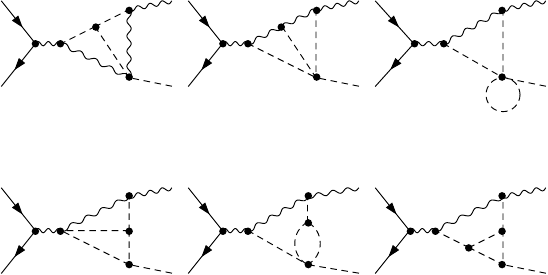}
\end{center}
\vspace{0mm} 
\caption{\label{fig:eeZh2loop} Example 1PI two-loop contributions to the process $e^+ e^- \to Zh$. The solid, dashed, and wiggly lines represent electrons, the Higgs, and the $Z$ boson, respectively. The individual Feynman diagrams scale as $\kappa_3$, $\kappa_4$, $\kappa_5$, $\kappa_3^2$, $\kappa_3 \hspace{0.25mm} \kappa_4$, and $\kappa_3^3$. See main text for additional explanations.}
\end{figure}

The $\kappa_3$, $\kappa_4$, and $\kappa_5$ corrections to any on-shell Higgs production process can be divided into two categories. The first category consists of universal corrections that stem from WFR of the external Higgs fields. This contribution acts as a renormalization factor that applies uniformly to all vertices where the Higgs interacts with vector bosons or fermions. Referring to a generic amplitude for the production of $n$ Higgs bosons as $ {\cal A}_n$, the finite corrections associated with Higgs~WFR can be accounted for via (cf.~also~\cite{Senaha:2018xek,Braathen:2019pxr,Braathen:2019zoh,Braathen:2020vwo} in a different context)
\beq \label{eq:WFRcorrection}
\left ( \delta {\cal A}_n \right )_{Z_h} = \left [ \left ( \displaystyle \frac{Z_h^{\rm OS}}{Z_h^{\overline{\rm MS}}} \right )^{n/2} - 1 \right ] {\cal A}_n \,, 
\eeq
where $Z_h^{\rm OS}$ and $Z_h^{\overline{\rm MS}}$ correspond to the Higgs~WFR in the on-shell and $\overline{\rm MS}$ scheme, respectively. The rescaling~(\ref{eq:WFRcorrection}) ensures that physical predictions are made using the finite, renormalized Higgs field. Expanding the square bracket in~(\ref{eq:WFRcorrection}) up to ${\cal O} (\lambda^2)$, we obtain 
\beq \label{eq:finite2loop}
\left( \frac{Z_h^{\rm OS}}{Z_h^{\overline{\rm MS}}} \right )^{n/2} - 1 = \frac{n}{2} \left \{ \frac{\lambda}{(4 \pi)^2} \hspace{0.25mm} \delta Z_{h, \rm fin}^{(1)} + \left ( \frac{\lambda}{(4 \pi)^2} \right )^2 \left [ \delta Z_{h, \rm fin}^{(2)} + \frac{n - 2}{4} \hspace{0.25mm} \left ( \delta Z_{h, \rm fin}^{(1)} \right )^2 \right ] \right \} \,, 
\eeq
with $\delta Z_{h, \rm fin}^{(1)}$ and $\delta Z_{h, \rm fin}^{(2)}$ denoting the expressions in~(\ref{eq:dZ1}) and~(\ref{eq:dZ2}), respectively, but without the $1/\bar \epsilon$ term. The $\delta Z_{h, \rm fin}^{(1)}$ and $\delta Z_{h, \rm fin}^{(2)}$ terms represent the finite one-loop and two-loop corrections necessary to ensure the proper normalization of the renormalized Higgs propagator, maintaining a residue of 1 in the full quantum theory. Furthermore, from~(\ref{eq:dZ1}) and~(\ref{eq:dZ2}), it follows that~(\ref{eq:finite2loop}) includes two-loop contributions proportional to $\kappa_4^2$, $\kappa_3 \hspace{0.25mm} \kappa_5$, $\kappa_3^3$, $\kappa_3^2 \hspace{0.25mm} \kappa_4$, and~$\kappa_3^4$. 

The second category of corrections is process dependent. For instance, in the case of $e^+ e^- \to Zh$ production, relevant 1PI diagrams that contribute at the two-loop level are depicted in~Figure~\ref{fig:eeZh2loop}. These individual Feynman diagrams scale as $\kappa_3$, $\kappa_4$, $\kappa_5$, $\kappa_3^2$, $\kappa_3 \hspace{0.25mm} \kappa_4$, and~$\kappa_3^3$, which collectively account for all the $\kappa_3$, $\kappa_4$, and $\kappa_5$ dependencies in the 1PI contributions to the $e^+ e^- \to Zh$ amplitude, up to the second order in perturbation theory. This implies that the two-loop contributions to the $e^+ e^- \to Zh$ process that are proportional to $\kappa_4^2$, $\kappa_3 \hspace{0.25mm} \kappa_5$, $\kappa_3^2 \hspace{0.25mm} \kappa_4$, and~$\kappa_3^4$ arise entirely from Higgs~WFR and are captured by the expression in~(\ref{eq:finite2loop}). The same reasoning applies to the $\kappa_3$, $\kappa_4$, and $\kappa_5$ dependence of the three-loop amplitude for $gg \to h$ production. Notably, this implies that for single-Higgs production and decay processes, the two-loop Higgs~WFR constant plays the same role for the $\kappa_4$ corrections as the one-loop WFR constant does for the $\kappa_3$ corrections. A quadratic dependence on $\kappa_4$ first appears at the two-loop level, and is entirely due to the universal corrections associated with the Higgs~WFR.

\section{Numerics}
\label{sec:numerics}

We are now ready to analyze the numerical effects of the Higgs~WFR corrections derived in~(\ref{eq:finite2loop}). We begin by examining the changes in the inclusive single-Higgs production cross sections $\sigma_i$ for the channel $i$ up to ${\cal O} (\lambda^2)$. By including both the process-dependent and universal one-loop corrections of ${\cal O}(\lambda)$, along with the ${\cal O}(\lambda^2)$ contributions from Higgs~WFR, we obtain: 
\beq \label{eq:deltasigmai}
\begin{split} 
\delta \sigma_i & \simeq C_1^{\sigma_i} \left ( \kappa_3 - 1 \right ) - 1.536 \cdot 10^{-3} \left ( \kappa_3^2 - 1 \right ) + 1.498 \cdot 10^{-6} \left ( 1 - 1.333 \hspace{0.25mm} L \right ) \left ( \kappa_4^2 - 1 \right ) \\[2mm]
& \phantom{xx} -4.180 \cdot 10^{-7} \left ( 1 + L \right ) \kappa_3 \hspace{0.25mm} \kappa_5 - 8.517 \cdot 10^{-6} \left ( 1 - 1.767 L \right ) \left ( \kappa_3^2 \hspace{0.25mm} \kappa_4 - 1 \right ) \\[2mm]
& \phantom{xx} - 1.726 \cdot 10^{-5} \left ( 1 + 1.307 L \right ) \left ( \kappa_3^4 - 1 \right ) \,.
\end{split}
\eeq 
Here, we used $\lambda \simeq 0.1289$ and $C_1^{\sigma_i}$ represents the process-dependent relative corrections linear in $\kappa_3 - 1$, calculated for many Higgs production channels and different colliders in~\cite{Degrassi:2016wml,Bizon:2016wgr,Maltoni:2017ims,DiVita:2017eyz,DiVita:2017vrr,Maltoni:2018ttu,Gorbahn:2019lwq,Degrassi:2019yix,Durieux:2017rsg,Haisch:2024nzv}. We note that~(\ref{eq:deltasigmai}) does not include ${\cal O} (\lambda^2)$ corrections proportional to $\kappa_3^3 - 1$. As discussed in Section~\ref{sec:calculation}, this omission is due to the process-dependent nature of such corrections, unlike the ${\cal O} (\lambda^2)$ terms in~(\ref{eq:deltasigmai}). Including the $\kappa_3^3 - 1$ contribution would have a negligible effect, given its small coefficient in~(\ref{eq:dZ2}) relative to the other contributions.

In contrast to the production cross sections, the modifications to the Higgs branching ratios are unaffected by Higgs~WFR. We obtain
\beq \label{eq:deltaBRf}
\delta {\rm BR}_ f \simeq \frac{\left ( \kappa_3 - 1 \right ) \left ( C_1^{\Gamma_f} - C_1^{\Gamma_h} \right )}{1 + \left ( \kappa_3 - 1 \right ) C_1^{\Gamma_h}} \,,
\eeq
with~$C_1^{\Gamma_f}$ representing the process-dependent relative correction linear in $\kappa_3 - 1$ for the Higgs decay width into the final state $f$, and $C_1^{\Gamma_{h}} = 0.23 \hspace{.25mm} \cdot \hspace{.25mm} 10^{-2}$~\cite{Degrassi:2016wml,Bizon:2016wgr,Gorbahn:2019lwq} corresponds to the coefficient of the one-loop correction proportional to $\kappa_3 - 1$ in the total Higgs decay width relative to the SM value. In terms of~(\ref{eq:deltasigmai}) and~(\ref{eq:deltaBRf}), the Higgs signal strengths for production in the channel~$i$ and decay in the channel~$f$ can be written~as
\beq \label{eq:muif}
\mu_i^f \simeq 1 + \delta \sigma_i + \delta {\rm BR}_ f \,.
\eeq

To constrain $\kappa_3$, $\kappa_4$, and $\kappa_5$ from future precision measurements of single-Higgs production processes, we consider the following $\chi^2$ function
\beq \label{eq:chi2}
\chi^2 = \sum_{i,f}\frac{\big (\mu_i^f -1 \big )^2}{\big (\Delta^f_i \big )^2} \,,
\eeq
where $\mu_i^f$ has been defined in~(\ref{eq:muif}), and we have assumed that the central values of the future measurements of the Higgs signal strengths will coincide with the corresponding SM predictions. The variables $\Delta^f_i$ encode the relative total uncertainties obtained by combining the theoretical and statistical uncertainties associated with $\mu_i^f$. The allowed confidence limit~(CL) regions are then obtained by minimizing~(\ref{eq:chi2}) and determining the solutions to $\Delta \chi^2 = \chi^2 - \chi^2_{\rm min} < 2 \hspace{0.25mm} Q^{-1} (n/2, 1- {\rm CL})$ for a $n$ parameter fit, where $Q^{-1} (a, z)$ is the regularised incomplete~gamma~function.

\begin{table}[t!]
\begin{center}
\begin{tabular}{| c | c | c | c | c | c |} 
\hline 
& ggF & $Wh$ & $Zh$ & VBF & $t\bar{t}h$ \\
\hline 
$C_1^{\sigma_i}$ & $0.68 \cdot 10^{-2}$ & $1.03 \cdot 10^{-2}$ & $1.18 \cdot 10^{-2}$ & $0.64 \cdot 10^{-2}$ & $3.47 \cdot 10^{-2}$ \\
\hline \hline 
& $\gamma \gamma$ & $W^+W^-$ & $ZZ$ & $b \bar b$ & $\tau^+ \tau^-$ \\
\hline 
$C_1^{\Gamma_f}$ & $0.49 \cdot 10^{-2}$ & $0.73 \cdot 10^{-2}$ & $0.83 \cdot 10^{-2}$ & $0.67 \cdot 10^{-5}$ & $0.33 \cdot 10^{-5}$ \\
\hline 
\end{tabular}
\end{center}
\vspace{0mm} 
\caption{\label{tab:C1HLLHC} Values of the process-dependent coefficients $C_1^{\sigma_i}$ and $C_1^{\Gamma_f}$. The numbers are directly taken or obtained from~\cite{Degrassi:2016wml,Bizon:2016wgr,Gorbahn:2019lwq,Haisch:2024nzv}, with the coefficients $C_1^{\sigma_i}$ corresponding to $pp$ collisions at $\sqrt{s} = 14 \, {\rm TeV}$.}
\end{table}

In our numerical analysis of the HL-LHC reach for single-Higgs production processes, we include Higgs production via gluon-gluon-fusion~($\rm ggF$), in association with a $W$ or $Z$ boson~($Wh$, $Zh$), in vector-boson-fusion~($\rm VBF$), and in association with top quarks~($t \bar t h$). Regarding Higgs branching ratios, we consider decays to pairs of photons~($\gamma \gamma$), EW gauge bosons~($W^+ W^-$, $ZZ$), bottom quarks~($b \bar b$), and tau leptons~($\tau^+ \tau^-$). The associated coefficients~$C_1^{\sigma_i}$~and~$C_1^{\Gamma_f}$ are collected in Table~\ref{tab:C1HLLHC}. The coefficients $C_1^{\sigma_i}$ correspond to proton-proton ($pp$) collisions at a centre-of-mass energy of $\sqrt{s} = 14 \, {\rm TeV}$. The values of the relative total uncertainties~$\Delta^f_i$~used in our HL-LHC analysis are listed in Table~\ref{tab:Deltaif}. The quoted uncertainties align with the baseline scenario S2, as described in the ATLAS note~\cite{ATL-PHYS-PUB-2018-054}. This scenario assumes that all theoretical uncertainties are halved relative to our current understanding of the relevant signals and backgrounds, reflecting the anticipated situation at the end of the HL-LHC program.

\begin{table}[t!]
\begin{center}
\begin{tabular}{|c | c | c |} 
\hline 
production, decay & $\Delta_i^f$ \\
\hline \hline 
ggF, $h \to \gamma\gamma$ & 3.6\% \\
\hline 
ggF, $h \to W^+ W^-$ & 4.4\% \\
\hline 
ggF, $h \to ZZ$ & 3.9\% \\
\hline 
$Wh$, $h \to \gamma \gamma$ &13.8\% \\
\hline 
$Wh$, $h \to b \bar b$ &10.0\% \\
\hline 
$Zh$, $h \to \gamma \gamma$ & 15.7\% \\
\hline 
$Zh$, $h \to b \bar b$ & 5.2\% \\
\hline 
$Vh$, $h \to ZZ$ & 18.2\% \\
\hline 
VBF, $h \to \gamma \gamma$ & 8.9\% \\
\hline 
VBF, $h \to W^+ W^-$ & 6.6\% \\
\hline 
VBF, $h \to Z Z$ & 11.8\% \\
\hline 
VBF, $h \to \tau^+ \tau^-$ & 7.8\% \\
\hline 
$t \bar t h$, $h \to \gamma \gamma$ & 7.4\% \\
\hline 
$t \bar t h$, $h \to ZZ$ &19.3\% \\
\hline 
\end{tabular}
\end{center}
\vspace{0mm} 
\caption{\label{tab:Deltaif} Relative total uncertainties $\Delta^f_i$ on the Higgs signal strengths, as defined in~(\ref{eq:muif}). The~numbers are taken from the HL-LHC study~\cite{ATL-PHYS-PUB-2018-054}. They assume $\sqrt{s} = 14 \, {\rm TeV}$ and an integrated luminosity of~$3 \, {\rm ab}^{-1}$. The $Vh$ production channel represents a combination of the $Wh$ and $Zh$ channels. Additional details can be found in the main text.}
\end{table}

To gain a comprehensive understanding of the potential constraints on modifications to the cubic, quartic, and quintic Higgs self-couplings by the end of the HL-LHC, we also examine double-Higgs and triple-Higgs production. For the corresponding inclusive signal strengths in $pp$ collisions at $\sqrt{s} = 14 \, {\rm TeV}$, we obtain the following expressions:
\beq \label{eq:multiHLLHC}
\begin{split}
\mu_{2h}^{\text{HL-LHC}} & = 2.18 - 1.50 \hspace{0.25mm} \kappa_3 - 1.00 \cdot 10^{-3} \hspace{0.25mm} \kappa_4 + 3.24 \cdot 10^{-1} \hspace{0.25mm} \kappa_3^2 + 3.95 \cdot 10^{-3} \hspace{0.25mm} \kappa_3 \hspace{0.25mm} \kappa_4 \\[2mm]
& \phantom{xx} + 5.54 \cdot 10^{-5} \hspace{0.25mm} \kappa_4^2 - 1.56 \cdot 10^{-3} \hspace{0.25mm} \kappa_3^2 \hspace{0.25mm} \kappa_4 - 3.78 \cdot 10^{-5} \hspace{0.25mm} \kappa_3 \hspace{0.25mm} \kappa_4^2 + 9.56 \cdot 10^{-6} \hspace{0.25mm} \kappa_3^2 \hspace{0.25mm} \kappa_4^2 \,, \\[4mm]
\mu_{3h}^{\text{HL-LHC}} & = 2.97 - 3.31 \hspace{0.25mm} \kappa_3 + 9.17 \cdot 10^{-2} \hspace{0.25mm} \kappa_4 + 1.83 \hspace{0.25mm} \kappa_3^2 - 2.66 \cdot 10^{-1} \hspace{0.25mm} \kappa_3 \hspace{0.25mm} \kappa_4 \\[2mm] 
& \phantom{xx} + 1.71 \cdot 10^{-2} \hspace{0.25mm} \kappa_4^2 - 4.24 \cdot 10^{-1} \hspace{0.25mm} \kappa_3^3 + 4.91 \cdot 10^{-2} \hspace{0.25mm} \kappa_3^2 \hspace{0.25mm} \kappa_4 + 4.13 \cdot 10^{-2} \hspace{0.25mm} \kappa_3^4 \,.
\end{split}
\eeq 
We recall that the expression for $\mu_{2h}^{\text{HL-LHC}}$ depends on the choice of the quintic Higgs self-coupling. Following~\cite{Bizon:2018syu}, we have assumed $\kappa_5 = 3/4 \hspace{0.5mm} \bar c_6 + 7/2 \hspace{0.5mm} \bar c_8 = 7/4 - 9/4 \hspace{0.5mm} \kappa_3 + 1/2 \hspace{0.5mm} \kappa_4$. These relations follow from~(\ref{eq:kappa3kappa4kappa5}). The~formula for the signal strength in double-Higgs production was obtained using the \texttt{POWHEG~BOX}~\cite{Alioli:2010xd} implementation of $pp \to hh$ production, as described in~\cite{Bizon:2024juq}. This~implementation includes the relevant EW two-loop amplitudes involving a modified quartic and quintic Higgs self-interactions, as calculated in~\cite{Bizon:2018syu}, and supplements these effects with next-to-leading order (NLO) QCD corrections, incorporating variations of the cubic Higgs self-coupling with full top-quark mass dependence~\cite{Borowka:2016ehy,Borowka:2016ypz,Heinrich:2017kxx,Heinrich:2019bkc}. The~numerical coefficients in~$\mu_{2h}^{\text{HL-LHC}}$ correspond to the \texttt{PDF4LHC15\_nlo} parton distribution functions (PDFs)~\cite{Butterworth:2015oua}, using the scale choice~$\mu_R = \mu_F = m_{hh}/2$, where $\mu_R$ and $\mu_F$ denote the renormalization and factorization scales, respectively, and $m_{hh}$ is the invariant mass of the double-Higgs system. The expression for $\mu_{3h}^{\text{HL-LHC}}$, on the other hand, was obtained at leading order in QCD using \texttt{MadGraph5\_aMC@NLO}~\cite{Alwall:2014hca}, incorporating the NLO QCD corrections from~\cite{Maltoni:2014eza} as an overall normalization. The same PDF set, \texttt{PDF4LHC15\_nlo}, was~used. 

\begin{figure}[t!]
\begin{center}
\includegraphics[height=0.45\textwidth]{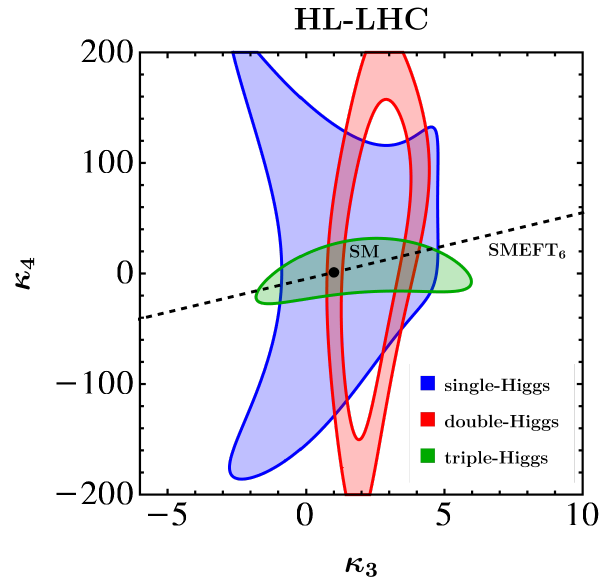} \qquad 
\includegraphics[height=0.45\textwidth]{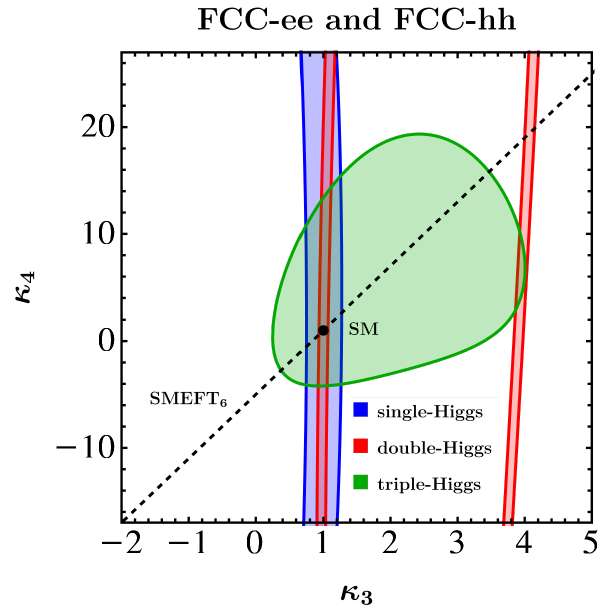}
\end{center}
\vspace{0mm} 
\caption{\label{fig:reach} Possible future constraints in the $\kappa_3\hspace{0.25mm}$--$\hspace{0.25mm}\kappa_4$~plane at the~HL-LHC~(left~panel) and the~FCC~(right~panel). The blue, red, and green contours correspond to the preferred 68\%~CL regions that arise from inclusive single-Higgs, double-Higgs, and triple-Higgs production, respectively. The SM is indicated by the black point, and the black dashed line corresponds to $\kappa_4 - 1 = 6 \left ( \kappa_3 - 1 \right )$,~i.e.,~the relation between $\kappa_3$ and $\kappa_4$ that holds in the SMEFT at the level of dimension-six operators. Consult~the main text for further explanations.}
\end{figure}

The results of our HL-LHC sensitivity study are displayed in the left panel of Figure~\ref{fig:reach}. The exclusions shown assume $pp$ collisions at $\sqrt{s} = 14 \, {\rm TeV}$ and an integrated luminosity of~$3 \, {\rm ab}^{-1}$. The blue, red, and green curves illustrate the hypothetical 68\%~CL limits from measurements of single-Higgs, double-Higgs, and triple-Higgs production, respectively. The~SM~point is indicated by the black dot, and the dashed black line corresponds to the family of solutions $\kappa_4 - 1 = 6 \left ( \kappa_3 - 1 \right )$. To derive the constraints from single-Higgs production, we followed the analysis strategy outlined above, fixing the renormalization scale appearing in~(\ref{eq:deltasigmai}) to $\mu = m_h$, and setting $\kappa_5 = 3/4 \hspace{0.5mm} \bar c_6 + 7/2 \hspace{0.5mm} \bar c_8 = 7/4 - 9/4 \hspace{0.5mm} \kappa_3 + 1/2 \hspace{0.5mm} \kappa_4$ in accordance with~(\ref{eq:multiHLLHC}). For double-Higgs and triple-Higgs production, we have assumed that at the HL-LHC, the corresponding signal strengths can be constrained to $0.77 < \mu_{2h}^{\text{HL-LHC}} < 1.23$ and $\mu_{3h}^{\text{HL-LHC}} < 10$. These hypothetical limits are consistent with those derived in~\cite{ATL-PHYS-PUB-2022-005} and~\cite{Stylianou:2023xit,Papaefstathiou:2023uum,Brigljevic:2024vuv,Dong:2025lkm}, respectively.

From the results depicted in the left panel of~Figure~\ref{fig:reach}, it is evident that at the HL-LHC, the new constraint arising from single-Higgs production is notably weaker than those stemming from double-Higgs and triple-Higgs production. As shown in~Appendix~\ref{app:moresingle}, this finding is independent of the specific choice of renormalization scale used in the single-Higgs production analysis. The same holds for the choice of the quintic Higgs self-coupling modifier $\kappa_5$, as we have explicitly verified, given its very small numerical coefficient in~(\ref{eq:deltasigmai}). After combining all Higgs measurements, we observe that at the HL-LHC, two regions of parameter space can be expected to remain viable, centered around the SM point and $\{ \kappa_3, \kappa_4 \} \simeq \{ 3.5, 0 \}$, respectively. In the case of $\kappa_3 = 1$, we find that at the HL-LHC it might be possible to obtain the following 68\%~CL constraint: $-21 < \kappa_4 < 28$. Moreover, the family of solutions $\kappa_4 - 1 = 6 \left( \kappa_3 - 1 \right)$ traverses a large portion of the non-SM region in the $\kappa_3\hspace{0.25mm}$--$\hspace{0.25mm}\kappa_4$ plane. This indicates that, using HL-LHC data, it will not be feasible to differentiate between BSM scenarios where significant modifications to the cubic, quartic, and quintic Higgs self-interactions stem solely from the operator $Q_6$, or from a combination of $Q_6$ and $Q_8$ --- cf.~the discussion following~(\ref{eq:kappa3kappa4kappa5}).

Our numerical analysis of the FCC reach for single-Higgs production processes focuses on the precision measurements that the lepton option of the FCC~(FCC-ee) is expected to achieve in the Higgsstrahlungs and VBF processes, i.e., $e^+ e^- \to Zh$ and $e^+ e^- \to \nu \bar \nu h$. The associated values of the process-dependent coefficients $C_1^{\sigma_i}$ and the relative total uncertainties $\Delta^f_i$ are collected in~Table~\ref{tab:C1DeltaifFCC}. The quoted numbers are taken from \cite{Durieux:2017rsg} and correspond to $e^+e^-$ collisions at $\sqrt{s} = 240 \, {\rm GeV}$ and an integrated luminosity of $10 \, {\rm ab}^{-1}$. Measurements of single-Higgs signal strengths at the hadron option of the FCC~(FCC-hh) are expected to reach a precision at the percent level \cite{FCC:2018byv}. This accuracy is not sufficient to compete with the few permille precision that the FCC-ee is likely to achieve in the $e^+ e^- \to Zh$ and $e^+ e^- \to \nu \bar \nu h$ channels, and we thus do not include hypothetical FCC-hh measurements in our FCC fit of single-Higgs observables.

Similar to the case of the HL-LHC, we also examine FCC-hh prospects for double-Higgs and triple-Higgs production to constrain the parameter space in the $\kappa_3\hspace{0.25mm}$--$\hspace{0.25mm}\kappa_4$~plane. Proceeding as described in the paragraph below~(\ref{eq:multiHLLHC}), we obtain the following expressions
\beq \label{eq:multiFCC}
\begin{split}
\mu_{2h}^{\text{FCC-hh}} & = 2.00 - 1.25 \hspace{0.25mm} \kappa_3 + 9.51 \cdot 10^{-6} \hspace{0.25mm} \kappa_4 + 2.56 \cdot 10^{-1} \hspace{0.25mm} \kappa_3^2 + 3.10 \cdot 10^{-3} \hspace{0.25mm} \kappa_3 \hspace{0.25mm} \kappa_4 \\[2mm]
& \phantom{xx} + 5.99 \cdot 10^{-5} \hspace{0.25mm} \kappa_4^2 - 1.23 \cdot 10^{-3} \hspace{0.25mm} \kappa_3^2 \hspace{0.25mm} \kappa_4 - 3.78 \cdot 10^{-5} \hspace{0.25mm} \kappa_3 \hspace{0.25mm} \kappa_4^2 + 8.70 \cdot 10^{-6} \hspace{0.25mm} \kappa_3^2 \hspace{0.25mm} \kappa_4^2 \,, \\[4mm]
\mu_{3h}^{\text{FCC-hh}} & = 2.57 - 2.62 \hspace{0.25mm} \kappa_3 + 3.91 \cdot 10^{-2} \hspace{0.25mm} \kappa_4 + 1.46 \hspace{0.25mm} \kappa_3^2 - 2.22 \cdot 10^{-1} \hspace{0.25mm} \kappa_3 \hspace{0.25mm} \kappa_4 \\[2mm] 
& \phantom{xx} + 1.55 \cdot 10^{-2} \hspace{0.25mm} \kappa_4^2 - 3.20 \cdot 10^{-1} \hspace{0.25mm} \kappa_3^3 + 3.99 \cdot 10^{-2} \hspace{0.25mm} \kappa_3^2 \hspace{0.25mm} \kappa_4 + 3.00 \cdot 10^{-2} \hspace{0.25mm} \kappa_3^4 \,,
\end{split}
\eeq 
for the signal strengths in double-Higgs and triple-Higgs production at the FCC-hh colliding protons at $\sqrt{s} = 100 \, {\rm TeV}$. In the case of $\mu_{2h}^{\text{FCC-hh}}$, we have again used~(\ref{eq:kappa3kappa4kappa5}) to fix the quintic Higgs self-coupling,~i.e., $\kappa_5 = 3/4 \hspace{0.5mm} \bar c_6 + 7/2 \hspace{0.5mm} \bar c_8 = 7/4 - 9/4 \hspace{0.5mm} \kappa_3 + 1/2 \hspace{0.5mm} \kappa_4$. This choice has also been made in~\cite{Bizon:2018syu,Bizon:2024juq}.

\begin{table}[t!]
\begin{center}
\begin{tabular}{| c | c | c || c | c | c | c |} 
\hline 
& $Zh$ & $\nu \bar \nu h$ & & $Zh, h \to b \bar b$ & $\nu \bar \nu h, h \to b \bar b$ \\
\hline 
$C_1^{\sigma_i}$ & $0.17 \cdot 10^{-1}$ & $0.64 \cdot 10^{-2}$ & $\Delta_i^f$ & 0.20\% & 0.28\% \\
\hline
\end{tabular}
\end{center}
\vspace{-2mm} 
\caption{\label{tab:C1DeltaifFCC} Values of the process-dependent coefficients $C_1^{\sigma_i}$ and the corresponding relative total uncertainties $\Delta^f_i$ on the Higgs signal strengths. The numbers are taken from~\cite{Durieux:2017rsg} and correspond to electron-electron ($e^+e^-$) collisions at $\sqrt{s} = 240 \, {\rm GeV}$ and assume an integrated luminosity of~$10 \, {\rm ab}^{-1}$.}
\end{table}

The blue, red, and green contours in the right panel of Figure~\ref{fig:reach} illustrate the 68\%~CL bounds obtained from hypothetical FCC measurements of single-Higgs, double-Higgs, and triple-Higgs production, respectively. The black dot indicates the SM point, while the dashed black line corresponds to the region of the parameter space where $\kappa_4 - 1 = 6 \left( \kappa_3 - 1 \right)$. The single-Higgs limits arise from precision FCC-ee measurements of $e^+ e^- \to Zh$ and $e^+ e^- \to \nu \bar \nu h$, as described earlier, with the renormalization scale in~(\ref{eq:deltasigmai}) set to $\mu = m_h$. In accordance with~(\ref{eq:multiFCC}), we have also set $\kappa_5 = 3/4 \hspace{0.5mm} \bar c_6 + 7/2 \hspace{0.5mm} \bar c_8 = 7/4 - 9/4 \hspace{0.5mm} \kappa_3 + 1/2 \hspace{0.5mm} \kappa_4$. The constraints from double-Higgs and triple-Higgs production are based on $30 \, {\rm ab}^{-1}$ of $\sqrt{s} = 100 \, {\rm TeV}$ data collected at the FCC-hh. At such a future hadron collider, it is expected that the double-Higgs and triple-Higgs signal strengths can be bounded to the ranges $0.95 < \mu_{2h}^{\text{FCC-hh}} < 1.05$~\cite{Barr:2014sga,Azatov:2015oxa,He:2015spf,Contino:2016spe,Mangano:2016jyj,Goncalves:2018qas,Chang:2018uwu} and $\mu_{3h}^{\text{FCC-hh}} < 2$~\cite{Contino:2016spe,Mangano:2016jyj,Papaefstathiou:2015paa, Chen:2015gva,Fuks:2015hna,Kilian:2017nio,Fuks:2017zkg}, respectively.

The right panel of~Figure~\ref{fig:reach} shows the results of our sensitivity study at the FCC. We~observe that, compared to the HL-LHC, the constraints arising from single-Higgs production are significantly improved relative to the multi-Higgs production bounds. This improvement is due to the exceptional precision of a few permille that the FCC-ee is expected to achieve in measuring the $e^+ e^- \to Zh$ and $e^+ e^- \to \nu \bar \nu h$ production cross sections. Additionally, by combining single-Higgs, double-Higgs, and triple-Higgs production, the FCC should be able to rule out parameter choices in the $\kappa_3\hspace{0.25mm}$--$\hspace{0.25mm}\kappa_4$ plane close to $\{ \kappa_3, \kappa_4 \} \simeq \{ 3.5, 0 \}$. Large modifications of the quartic Higgs self-coupling could then arise only from the simultaneous presence of the operators $Q_6$ and $Q_8$. It is important to emphasize that incorporating single-Higgs production plays a key role in achieving this constraint, as clearly illustrated by the plot. Furthermore, the allowed region around the SM point will be significantly reduced at the FCC compared to the HL-LHC. Assuming $\kappa_3 = 1$, we obtain the limit $-7 < \kappa_4 < 15$. As before, these findings are largely independent of the specific choice of renormalization scale~$\mu$ and the quintic Higgs self-coupling modifier $\kappa_5$ in the single-Higgs analysis --- cf.~Appendix~\ref{app:moresingle} for further details. 

\section{Conclusions}
\label{sec:conclusions}

In this article, we have studied the possibility of constraining quartic Higgs self-interactions indirectly through future precision measurements of single-Higgs production processes. To~this~end, we have calculated the corrections to the Higgs~WFR constant up to the two-loop level that arise from modified cubic, quartic, and quintic Higgs self-couplings. By~employing integration-by-parts (IBP) identities, we have reduced the relevant two-loop Higgs self-energy diagrams to a set of scalar master integrals. All master integrals can be obtained in analytic form. The explicit expressions for these two-loop integrals are collected in~Appendix~\ref{app:integrals}. 

Utilizing our analytic result for the two-loop Higgs~WFR constant~(\ref{eq:dZ2}), we have performed an exploratory study of the two-dimensional constraints in the $\kappa_3 \hspace{0.25mm}$--$ \hspace{0.25mm} \kappa_4$ plane that precision measurements of single-Higgs production at the HL-LHC and the FCC-ee may allow us to set --- in light of the recent ATLAS results on triple-Higgs production~\cite{ATLAS:2024xcs}, we also provide an update on the current status of Higgs self-coupling determinations after LHC~Run~2 in Appendix~\ref{app:LHCRun2}. Our novel constraints have been contrasted with those that searches for multi-Higgs production might provide at the same facilities. In all cases, we have considered only measurements of total rates. Our analysis has shown that at the HL-LHC, the indirect limits on the quartic Higgs self-coupling derived from single-Higgs production are generically very weak, allowing for two orders of magnitude enhancements of $\left| \kappa_4 \right|$. This is due to the limited precision of single-Higgs measurements at the HL-LHC, which is restricted to a few percent. The leading two-dimensional limits on $\kappa_3$ and $\kappa_4$ at the HL-LHC will therefore arise from a combination of measurements of double-Higgs and triple-Higgs production. Combining all Higgs measurements, we observe that at the HL-LHC, two regions of parameter space will likely remain viable: one centered around the SM point and another located in the vicinity of $\{ \kappa_3, \kappa_4 \} \simeq \{ 3.5, 0 \}$. Focusing on the solution near the SM point, we find that $\left| \kappa_3 - 1 \right| \lesssim 30\%$ and $\left| \kappa_4 \right| \lesssim 20$.

Since FCC-ee measurements of Higgsstrahlungs and VBF production via $e^+ e^- \to Zh$ and $e^+ e^- \to \nu \bar{\nu} h$, respectively, are expected to reach a precision of a few permille, the indirect sensitivity of single-Higgs production to modifications of $\kappa_3$ and $\kappa_4$ is significantly enhanced at the FCC. In fact, our new indirect limits on the size of the cubic and quartic Higgs self-couplings are comparable in strength to those derived from double-Higgs production. The latter indirect limits have already been studied in detail in~\cite {Bizon:2018syu,Borowka:2018pxx}. A~combination of single-Higgs, double-Higgs, and triple-Higgs production therefore leads to notably improved prospects for the determination of the quartic Higgs self-coupling at the FCC. In~particular, a global analysis of Higgs observables in the FCC era is expected to be able to completely rule out the second BSM solution at $\{ \kappa_3, \kappa_4 \} \simeq \{ 3.5, 0 \}$. Importantly, single-Higgs production plays a pivotal role in reaching this conclusion. The FCC is expected to constrain deviations in $\kappa_3$ and $\kappa_4$ to the ranges $\left| \kappa_3 - 1 \right| \lesssim 10\%$ and $-5 \lesssim \kappa_4 \lesssim 15$. We have argued that the results from our single-Higgs analysis, both at the HL-LHC and FCC-ee, are largely independent of the specific choice of renormalization scale and the quintic Higgs self-coupling modifier $\kappa_5$. More details on the former point can be found in~Appendix~\ref{app:moresingle}. 

Other studies of the HL-LHC and FCC prospects for determining the quartic Higgs self-coupling can be found in~\cite{Liu:2018peg,Bizon:2018syu,Borowka:2018pxx,Stylianou:2023xit,Papaefstathiou:2023uum,Brigljevic:2024vuv,Dong:2025lkm}. The reach in bounding $\kappa_4$ at various lepton machines, including a muon collider, has been studied in~\cite{Maltoni:2018ttu,Gonzalez-Lopez:2020lpd,Chiesa:2020awd}. In all these works, the constraints on the quartic Higgs self-interactions follow from analyses of multi-Higgs~production. This article is the only one to date that also examines single-Higgs production~processes.

\acknowledgments UH thanks Christoph~Englert, Martin~Gorbahn, Luca~Rottoli, Chen-Yu~Wang, and Simone~Zoia for useful discussions and helpful correspondence. He also acknowledges computer support from Thomas~Hahn, Luc~Schnell, and Andrii~Verbytskyi. The Feynman diagrams shown in this work have been generated and drawn with \texttt{FeynArts}. Some of the MC simulations presented here were carried out using the Max Planck Computing and Data Facility (MPCDF) in Garching.

\newpage

\appendix

\section{Master integrals}
\label{app:integrals}

In this appendix, we provide the analytic results for all the master integrals that arise in the two-loop computation of the pure Higgs contribution to the Higgs self-energy. The scalar one-loop and two-loop Feynman integrals appearing in the calculation performed in this article are defined as follows
\bea \label{eq:MIsdef}
\begin{split}
A (0) & = N _d \int \frac{d^d l_1}{l_1^2 - m^2} \,, \\[2mm]
B (p^2) & = N _d \int \frac{d^d l_1}{\big (l_1^2 - m^2 \big) \big (\left (l_1 - p \right )^2 - m^2 \big )} \,, \\[2mm]
K (0) & = N _d ^2 \int \frac{d^d l_1 \hspace{0.25mm} d^d l_2}{\big (l_1^2 - m^2 \big) \big ( l_2^2 - m^2 \big ) \big ( ( l_1 - l_2 )^2 - m^2 \big )} \,, \\[2mm]
J (p^2) & = N _d ^2 \int \frac{d^d l_1 \hspace{0.25mm} d^d l_2}{\big (l_1^2 - m^2 \big) \big ( ( l_2 - p )^2 - m^2 \big ) \big ( ( l_1 - l_2 )^2 - m^2 \big )} \,, \\[2mm]
V (p^2) & = N _d ^2 \int \frac{d^d l_1 \hspace{0.25mm} d^d l_2}{\big (l_1^2 - m^2 \big) \big ( ( l_1 - p ) ^2 - m^2 \big ) \big ( ( l_2 - p )^2 - m^2 \big ) \big ( ( l_1 - l_2 )^2 - m^2 \big )} \,, \\[2mm]
F (p^2) & = N _d ^2 \int \frac{d^d l_1 \hspace{0.25mm} d^d l_2}{\big (l_1^2 - m^2 \big) \big (l_2^2 - m^2 \big) \big ( ( l_1 - p ) ^2 - m^2 \big ) \big ( ( l_2 - p )^2 - m^2 \big ) \big ( ( l_1 - l_2 )^2 - m^2 \big )} \,,
\end{split}
\eea
where $N _d = (\mu^2)^{(4-d)/2}/\pi^{d/2}$ is the normalization of the loop integrals. The above definitions resemble those used in~the publications~\cite{Tarasov:1997kx,Mertig:1998vk}. 

When evaluated for on-shell kinematics,~i.e.,~at~$p^2 = m^2$, we find the following analytic results
\beq \label{eq:MIs1}
\begin{split}
A (0) & = i \hspace{0.25mm} m^2 \left [ \frac{1}{\epsilon} + 1 + \left ( 1 + \frac{\zeta (2)}{2} \right) \epsilon \right ] N _\epsilon \,, \\[2mm]
B (m^2) & = i \hspace{0.25mm} \left \{ \frac{1}{\epsilon} + 2 - \frac{\pi}{\sqrt{3}} + \left [ 4 + \frac{\zeta (2)}{2} - \frac{\pi \left ( 2 - \ln 3 \right )}{\sqrt{3}} - \frac{4}{\sqrt{3}} \hspace{0.25mm} {\rm Cl}_2 \left ( \frac{\pi}{3} \right ) \right ] \epsilon \right \} \hspace{0.25mm} N _\epsilon \,, \\[2mm]
K (0) & = m^2 \left [ -\frac{3}{2 \hspace{0.25mm} \epsilon^2} -\frac{9}{2 \hspace{0.25mm} \epsilon} - \frac{21}{2} - \frac{3 \hspace{0.25mm} \zeta(2)}{2} + 2 \sqrt{3} \hspace{0.25mm} {\rm Cl}_2 \left ( \frac{\pi}{3} \right ) \right ] N _\epsilon^2 \,, \\[2mm] 
J (m^2) & = m^2 \left [ -\frac{3}{2 \hspace{0.25mm} \epsilon^2} -\frac{17}{4 \hspace{0.25mm} \epsilon} - \frac{59}{8} - \frac{3 \hspace{0.25mm} \zeta(2)}{2} \right ] N _\epsilon^2 \,, \\[2mm]
V (m^2) & = \left [ -\frac{1}{2 \hspace{0.25mm} \epsilon^2} - \frac{\frac{5}{2} - \frac{\pi}{\sqrt{3}}}{\epsilon} - \frac{19}{2} + \frac{\pi \left ( 4 - \ln 3 \right )}{\sqrt{3}} + \frac{7}{\sqrt{3}} \hspace{0.25mm} {\rm Cl}_2 \left ( \frac{\pi}{3} \right ) \right ] N _\epsilon^2 \,, \\[2mm]
F (m^2) & = \frac{1}{m^2} \left [ -\zeta(3) + \frac{2 \hspace{0.25mm} \pi}{3} \hspace{0.25mm} {\rm Cl}_2 \left ( \frac{\pi}{3} \right ) \right ] N _\epsilon^2 \,, \\[2mm]
\end{split}
\eeq
for the $\epsilon$ expansion of the master integrals defined in~(\ref{eq:MIsdef}). Here, $N _\epsilon = \left ( \mu^2/m^2 \hspace{0.5mm} e^{\gamma_E} \right )^\epsilon$. The~analytic expressions given in~(\ref{eq:MIs1}) agree with known results in the literature (cf.~for instance~\cite{Davydychev:1992mt,Scharf:1993ds,Fleischer:1998dw}). They have also been cross-checked against high-precision numerical results obtained from both the~\texttt{AMFlow}~\cite{Liu:2022chg} and \texttt{TSIL}~\cite{Martin:2005qm} packages. Our \texttt{AMFlow} installations rely on \texttt{LiteRed} and \texttt{FIRE}~\cite{Smirnov:2019qkx} for the IBP reductions of all intermediate integrals appearing in the calculation.

Besides the~one-loop and two-loop master integrals~(\ref{eq:MIsdef}), we also encounter in our computations the first derivatives of self-energy diagrams with respect to $p^2$ evaluated on-shell. Employing the notation introduced in~(\ref{eq:onshelldZ1}), the $\epsilon$ expansions for the relevant integrals take the following form: 
\beq \label{eq:MIs2}
\begin{split}
B^\prime (m^2) & = \frac{i}{m^2} \left \{ - 1 + \frac{2 \hspace{0.25mm} \pi}{3 \sqrt{3}} - \left [ 2 - \frac{\pi \left ( 3 - 2 \ln 3 \right )}{3\sqrt{3}} - \frac{8}{3 \sqrt{3}} \hspace{0.25mm} {\rm Cl}_2 \left ( \frac{\pi}{3} \right ) \right ] \epsilon \right \} \hspace{0.25mm} N _\epsilon \,, \\[2mm]
J^\prime (m^2) & = \left [ \frac{1}{4 \hspace{0.25mm} \epsilon} - \frac{3}{8} \right ] N _\epsilon^2 \,, \\[2mm]
V^\prime (m^2) & = \frac{1}{m^2} \left [ \frac{1 - \frac{2 \pi}{3 \sqrt{3}} }{\epsilon} + 5 - \frac{\pi \left ( 7 - 2 \ln 3 \right )}{3 \sqrt{3}} - \frac{\zeta (2)}{3} - \frac{8}{3 \sqrt{3}} \hspace{0.25mm} {\rm Cl}_2 \left ( \frac{\pi}{3} \right ) \right ] N _\epsilon^2 \,, \\[2mm]
F^\prime (m^2) & = \frac{1}{m^4} \left [ \frac{2 \zeta(2)}{3} + \zeta(3) - \frac{2 \pi}{3} \hspace{0.25mm} {\rm Cl}_2 \left ( \frac{\pi}{3} \right ) \right ] N _\epsilon^2 \,.
\end{split}
\eeq
The results for $B^\prime (m^2)$, $J^\prime (m^2)$, and the UV pole of $V^\prime (m^2)$ agree with known results. The~constant parts of $V^\prime (m^2)$ and $F^\prime (m^2)$, on the other hand, are, to our knowledge, new. The derivation was carried out using two methods. In the first approach, we utilized IBP identities generated with \texttt{LiteRed}. In the second approach, we employed the \texttt{PSLQ} algorithm~\cite{PSLQ} to analyze high-precision numerical results for $V^\prime (m^2)$ and $F^\prime (m^2)$, which were obtained using \texttt{AMFlow} and \texttt{TSIL}. Both methods yield identical final results.

\section{Counterterms}
\label{app:counterterms}

In this appendix, we list the Feynman rules for the counterterms that enter into our calculation. The counterterm for the Higgs propagator is given by
\beq \label{eq:higgsprop}
\raisebox{-2.5mm}{\includegraphics[width=0.25\textwidth]{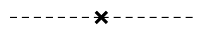}} \, = \; i \hspace{0.25mm} \Big [ \big ( p^2 - m_h^2 \big ) \hspace{0.25mm} \delta Z_h - \delta m_h^2 \Big ] \,, 
\eeq
while the counterterm for the cubic Higgs self-coupling takes the following form 
\beq \label{eq:higgstri}
\raisebox{-9.75mm}{\includegraphics[width=0.25\textwidth]{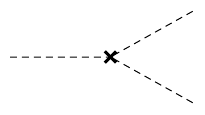}} \, = \; - 6 \hspace{0.25mm} i \hspace{0.25mm} \lambda \hspace{0.25mm} v \hspace{0.25mm} \kappa_3 \left ( \frac{\delta t}{v \hspace{0.25mm} m_h^2} + \frac{\delta m_h^2}{m_h^2} + \frac{3}{2} \hspace{0.25mm} \delta Z_h + \delta \lambda \right ) \,.
\eeq
Here 
\beq \label{eq:cts}
\begin{split}
\delta t & = -\frac{3 \hspace{0.25mm} \lambda \hspace{0.25mm} v \hspace{0.25mm} m_h^2}{(4 \pi)^2} \left ( \frac{1}{\bar \epsilon} + 1 + L \right ) \kappa_3 \,, \\[2mm]
\delta m_h^2 & = \frac{9 \hspace{0.25mm} \lambda \hspace{0.25mm} m_h^2}{(4 \pi)^2} \left ( \frac{1}{\bar \epsilon} + 2 - \frac{\pi}{\sqrt{3}} + L \right ) \kappa_3^2 + \frac{3 \hspace{0.25mm} \lambda \hspace{0.25mm} m_h^2}{(4 \pi)^2} \left ( \frac{1}{\bar \epsilon} + 1 + L \right ) \kappa_4 \,, \\[2mm]
\delta Z_h & = \frac{\lambda}{(4 \pi)^2} \left ( 9 - 2 \sqrt{3} \pi \right ) \kappa_3^2 \,, \\[2mm]
\delta \lambda & = \frac{1}{( 4 \pi )^2} \hspace{0.25mm} \frac{1}{\bar \epsilon} \left [ 3 \hspace{0.25mm} \lambda \left ( \kappa_3 + 2 \kappa_4 - 3 \kappa_3^2 \right ) - \frac{\kappa_5}{6 \hspace{0.125mm} \kappa_3} \right ] \,.
\end{split}
\eeq
Notice that the tadpole, mass, and WFR counterterms, i.e., $\delta t$, $\delta m_h^2$, and $\delta Z_h$, all correspond to the on-shell scheme. The $\delta \lambda$ counterterm, however, has been determined in the $\overline{\text{MS}}$ scheme by the requirement that it cancels the UV pole remaining in the cubic Higgs self-coupling after incorporating the counterterms $\delta t$, $\delta m_h^2$, and $\delta Z_h$. It also accounts for the additional UV poles that describe operator mixing in the SMEFT. Note that in the SM, i.e., for $\kappa_3 = \kappa_4 = 1$ and $\kappa_5 = 0$, we have $\delta \lambda = 0$, and the loop-corrected cubic Higgs self-coupling becomes UV finite after performing the tadpole, mass, and WFR corrections as expected.

\section{Scale dependence}
\label{app:moresingle}

\begin{figure}[t!]
\begin{center}
\includegraphics[height=0.45\textwidth]{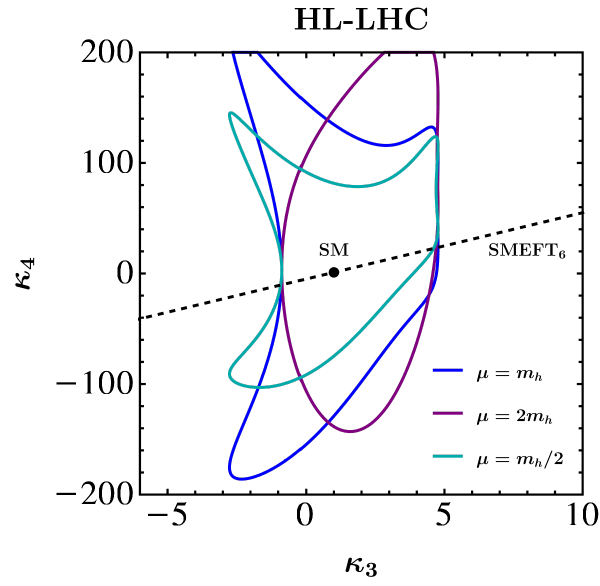} \qquad
\includegraphics[height=0.45\textwidth]{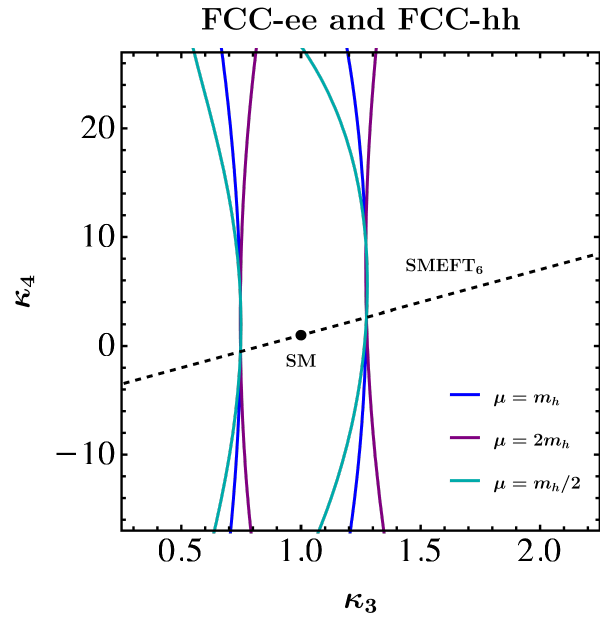} 
\end{center}
\vspace{0mm} 
\caption{\label{fig:moresingle} Single-Higgs production constraints in the $\kappa_3\hspace{0.25mm}$--$\hspace{0.25mm}\kappa_4$~plane at the~HL-LHC~(left~panel) and the~FCC~(right~panel). The blue, purple, and cyan contours correspond to the preferred 68\%~CL regions for the three choices of the renormalisation scale $\mu = m_h$, $\mu = 2 m_h$, and $\mu = m_h/2$, respectively. The other ingredients in the plots resemble those introduced in~Figure~\ref{fig:reach}.}
\end{figure}

In this appendix, we study the phenomenological impact of the renormalization scale dependence of the two-loop contribution~(\ref{eq:dZ2}) to the Higgs~WFR constant. To~this~end, we display the single-Higgs production constraints that follow from~(\ref{eq:deltasigmai}),~(\ref{eq:deltaBRf}), and~(\ref{eq:muif}) for three different choices of the renormalization scale. The outcome of this analysis is shown in~Figure~\ref{fig:moresingle}. The blue contours in both panels correspond to the choice $\mu = m_h$, which was also used to obtain the single-Higgs production limits displayed in~Figure~\ref{fig:reach}. The purple and cyan curves depict the bounds for $\mu = 2 m_h$ and $\mu = m_h/2$, respectively. All predictions correspond to the choice $\kappa_5 = 3/4 \hspace{0.5mm} \bar c_6 + 7/2 \hspace{0.5mm} \bar c_8 = 7/4 - 9/4 \hspace{0.5mm} \kappa_3 + 1/2 \hspace{0.5mm} \kappa_4$ of the quintic Higgs self-coupling. These relations follow from~(\ref{eq:kappa3kappa4kappa5}). From the results shown in the figure, it is clear that while the choice of renormalization scale impacts the obtained exclusions, it does not alter the overall picture. At the HL-LHC, the limits on the quartic Higgs self-coupling derived from single-Higgs production are generally very loose and not competitive with the bounds from double-Higgs and triple-Higgs production. In contrast, the scale dependence of the exclusions around the SM point is less pronounced at the FCC-ee, thanks to the anticipated high-precision measurements of $e^+ e^- \to Zh$ and $e^+ e^- \to \nu \bar \nu h$. Measurements of single-Higgs properties may therefore provide complementary and relatively model-independent information on $\kappa_3$ and $\kappa_4$ in the FCC era.

\section{LHC Run 2 results}
\label{app:LHCRun2}

In view of the recent ATLAS results on triple-Higgs production~\cite{ATLAS:2024xcs}, this appendix provides an overview of the current status of Higgs self-coupling extractions after LHC~Run~2. To~obtain the relevant constraints, we follow the methodology described in Section~\ref{sec:numerics}. For the single-Higgs production processes, our analysis is based on the measured signal strengths for the full set of simplified template cross sections provided by ATLAS in~\cite{ATLAS:2024lyh}. In the case of double-Higgs production, we use the upper limit on the production rate derived by ATLAS in~\cite{ATLAS:2024ish} from a combination of the searches for the $b \bar b b \bar b$~\cite{ATLAS:2023qzf,ATLAS:2024lsk}, $b \bar b \tau^+ \tau^-$~\cite{ATLAS:2024pov}, $b \bar b \gamma \gamma$~\cite{ATLAS:2023gzn}, and $b \bar b \ell^+ \ell^-\nu \bar \nu$~\cite{ATLAS:2023elc} final states.

\begin{figure}[t!]
\begin{center}
\includegraphics[height=0.45\textwidth]{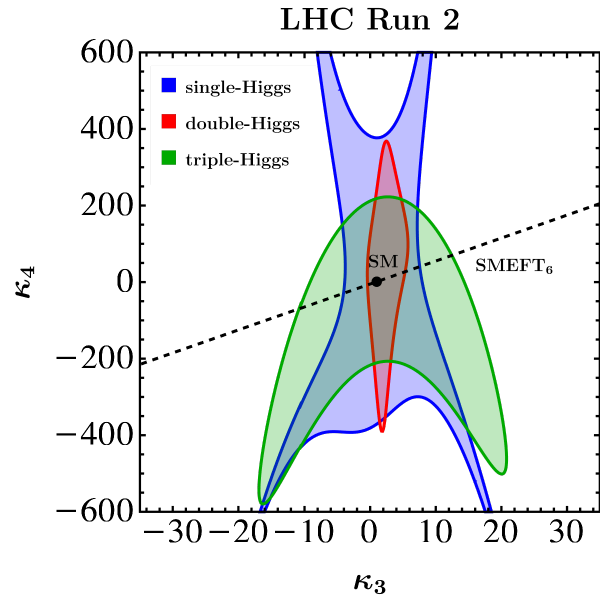} \qquad 
\includegraphics[height=0.45\textwidth]{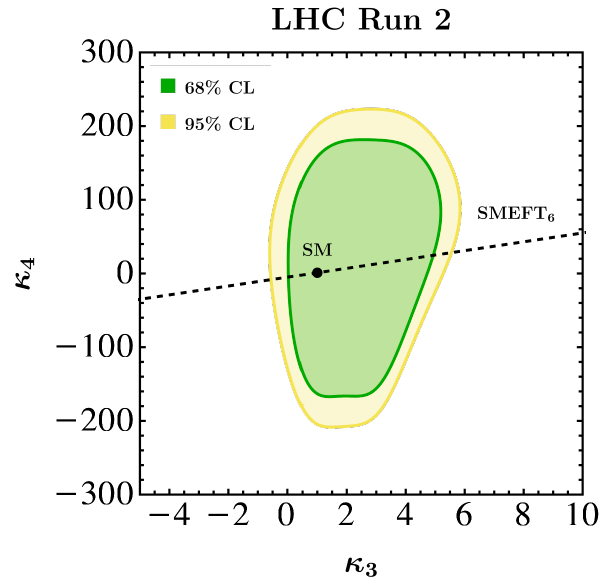} 
\end{center}
\vspace{-4mm} 
\caption{\label{fig:LHCRun2} Left panel: Constraints in the $\kappa_3\hspace{0.25mm}$--$\hspace{0.25mm}\kappa_4$~plane after LHC Run 2. The blue, red, and green contours indicate the 95\%~CL preferred regions based on inclusive single-Higgs, double-Higgs, and triple-Higgs production, respectively. Right panel: Combined constraints in the $\kappa_3\hspace{0.25mm}$--$\hspace{0.25mm}\kappa_4$~plane after LHC Run 2. The green and yellow contours correspond to the 68\%~CL and 95\%~CL regions, respectively. In both panels, the black points denote the SM prediction, and the black dashed lines represent the relation $\kappa_4 - 1 = 6 \left( \kappa_3 - 1 \right)$, which arises in the SMEFT framework at the dimension-six operator level. Additional details are provided in the main text.}
\end{figure}

The results of our LHC~Run~2~study are shown in~Figure~\ref{fig:LHCRun2}. The blue, red, and green contours in the left panel represent the 95\%~CL preferred regions obtained from inclusive single-Higgs, double-Higgs, and triple-Higgs production, respectively. The~SM~point is indicated by the black dot, and the dashed black line corresponds to the parameter space satisfying $\kappa_4 - 1 = 6 \left ( \kappa_3 - 1 \right )$. To derive the constraints from single-Higgs production, we followed the analysis strategy described in Section~\ref{sec:numerics}, fixing the renormalization scale to $\mu = m_h$, and setting $\kappa_5 = 3/4 \hspace{0.5mm} \bar c_6 + 7/2 \hspace{0.5mm} \bar c_8 = 7/4 - 9/4 \hspace{0.5mm} \kappa_3 + 1/2 \hspace{0.5mm} \kappa_4$. The double-Higgs and triple-Higgs constraints shown assume $\mu_{2h}^{\text{LHC~Run~2}} < 2.9$~\cite{ATLAS:2024ish} and $\mu_{3h}^{\text{LHC~Run~2}} < 760$~\cite{ATLAS:2024xcs}. Our calculation of the double-Higgs and triple-Higgs signal strengths employs the same Monte~Carlo codes used in Section~\ref{sec:numerics}. The plot clearly shows that the constraints from double-Higgs and triple-Higgs production are largely complementary. Specifically, $pp \to 2h$ offers the strongest constraint on $\kappa_3$ when $\kappa_4 = 1$, whereas $pp \to 3h$ provides the most stringent limit on $\kappa_4$ when $\kappa_3 = 1$. The constraint from single-Higgs production is significantly weaker compared to those from double-Higgs and triple-Higgs production. These findings are largely unaffected by the particular choice of renormalization scale in the single-Higgs production analysis, as well as by the choice of the quintic Higgs self-coupling modifier $\kappa_5$, which influences both our double-Higgs and single-Higgs predictions. In addition, the right panel of Figure~\ref{fig:LHCRun2} shows the combined constraints in the $\kappa_3\hspace{0.25mm}$--$\hspace{0.25mm}\kappa_4$ plane after LHC Run 2, with the green and yellow contours indicating the 68\%~CL and 95\%~CL regions, respectively. Assuming $\kappa_4 = 1$, our combined analysis constrains $\kappa_3$ to the 95\%~CL interval $-0.6 < \kappa_3 < 5.3$. Conversely, fixing $\kappa_3 = 1$ yields a corresponding 95\%~CL range of $-205 < \kappa_4 < 200$. These limits are in line with the bounds on $\kappa_3$ and $\kappa_4$ reported by ATLAS in~\cite{ATLAS:2022jtk} and~\cite{ATLAS:2024xcs}, respectively.


%

\end{document}